\documentclass{iopart}
\usepackage{amssymb} %
\usepackage{graphicx} %
\usepackage{color} %
\usepackage{bm} 
\usepackage{units} 
\usepackage{xspace}
\RequirePackage{ifpdf}

\newcommand{\eqref}[1]{{(\ref{#1})}}

\newcommand{\hyb}{\mathrm{hyb}}
\newcommand{\NR}{\mathrm{NR}}
\newcommand{\pN}{\mathrm{pN}}

\newcommand{\MSun}{\ensuremath{M_{\odot}}}
\newcommand{\deff}{\ensuremath{D_{\mathrm{eff}}}} 
\newcommand{\ta}{\ensuremath{t_{\mathrm{a}}}}
\newcommand{\phia}{\ensuremath{\phi_{\mathrm{a}}}}

\renewcommand{\etal}{\textit{et al.}\xspace}
\newcommand{\G}{G}
\renewcommand{\c}{c}
\newcommand{\lvert}{\ensuremath{|}}
\newcommand{\rvert}{\ensuremath{|}}

\makeatletter
\let\protect\relax
{\catcode`\|=\active
  \xdef\InnerProduct{\protect\expandafter\noexpand\csname InnerProduct \endcsname}
  \expandafter\gdef\csname InnerProduct \endcsname#1{%
    \begingroup
    \ifx\SavedDoubleVert\relax
    \let\SavedDoubleVert\|\let\|\IpDoubleVert
    \fi
    \mathcode`\|32768\let|\IPVert
    \left({#1}\right)
    \endgroup
  }
}
\def\IPVert{\@ifnextchar|{\|\@gobble}
     {\egroup\,\mid@vertical\,\bgroup}}
\def\IPDoubleVert{\egroup\,\mid@dblvertical\,\bgroup}
\let\SavedDoubleVert\relax
\def\midvert{\egroup\mid\bgroup}
\def\SetVert{\@ifnextchar|{\|\@gobble}
    {\egroup\;\mid@vertical\;\bgroup}}
\def\SetDoubleVert{\egroup\;\mid@dblvertical\;\bgroup}
\def\mid@vertical{\mskip1mu\vrule\mskip1mu}
\def\mid@dblvertical{\mskip1mu\vrule\mskip2.5mu\vrule\mskip1mu}
\makeatother
\usepackage{braket}
\newcommand{\Overlap}{\Braket}
\usepackage{ulem}
\begin{document}

\title[Black hole binaries: High-accuracy simulations and
\textit{TaylorF2} template efficiency]{Comparison of high-accuracy
  numerical simulations of black-hole binaries with stationary phase
  post-{N}ewtonian template waveforms for {I}nitial and {A}dvanced
  {LIGO}.}

\author{Michael Boyle,${}^{1,2}$ Duncan A Brown${}^{3}$ and Larne
  Pekowsky${}^{3}$}

\address{$^{1}$ Theoretical Astrophysics 103-33,
  California Institute of Technology, Pasadena, CA 91125} %
\address{$^{2}$ 604 Space Sciences Building, Cornell University,
  Ithaca, NY 14853} %
\address{$^{3}$ Department of Physics, Syracuse University, Syracuse,
  New York, 13244}

\date{\today}

\begin{abstract}
  We study the effectiveness of stationary-phase approximated
  post-Newtonian waveforms currently used by ground-based
  gravitational-wave detectors to search for the coalescence of binary
  black holes by comparing them to an accurate waveform obtained from
  numerical simulation of an equal-mass non-spinning binary black hole
  inspiral, merger and ringdown.  We perform this study for the
  Initial- and Advanced-LIGO detectors.  We find that overlaps between
  the templates and signal can be improved by integrating the match
  filter to higher frequencies than used currently.  We propose simple
  analytic frequency cutoffs for both Initial and Advanced LIGO, which
  achieve nearly optimal matches, and can easily be extended to
  unequal-mass, spinning systems.  We also find that templates that
  include terms in the phase evolution up to 3.5 pN order are nearly
  always better, and rarely significantly worse, than 2.0 pN templates
  currently in use.  For Initial LIGO we recommend a strategy using
  templates that include a recently introduced pseudo-4.0 pN term in
  the low-mass ($M \leq 35\,\MSun$) region, and 3.5 pN templates
  allowing unphysical values of the symmetric reduced mass $\eta$
  above this.  This strategy always achieves overlaps within 0.3\% of
  the optimum, for the data used here.  For Advanced LIGO we recommend
  a strategy using 3.5 pN templates up to $M=12\,\MSun$, 2.0 pN
  templates up to $M=21\,\MSun$, pseudo-4.0 pN templates up to
  $65\,\MSun$, and 3.5 pN templates with unphysical $\eta$ for higher
  masses.  This strategy always achieves overlaps within 0.7\% of the
  optimum for Advanced LIGO.
\end{abstract}


\maketitle

\section{Introduction}
\label{sec:Introduction} %

The coalescence of binary black holes is one the most promising
sources of gravitational waves for interferometric gravitational-wave
detectors, such as LIGO, Virgo and GEO600~\cite{thorne.k:1987}. The
first-generation LIGO detectors have achieved their design sensitivity
and recorded over one year of coincident data~\cite{Abbott:2007kva}.
This data, together with data from the Virgo detector, are currently
being searched for gravitational waves from compact binary
coalescence~\cite{Abbott:2003pj,Abbott:2005pe,Abbott:2005pf,%
  Abbott:2007xi,Abbott:2007ai,Abbott:2008}.  Upgrades to improve the sensitivity
of these detectors by a factor of two, and ultimately 10, are
underway.  Optimal searches using the enhanced detectors in 2009 will
be sensitive to black-hole coalescence out to hundreds of
megaparsecs~\cite{LIGOEnhancedLIGO}. The advanced detectors,
operational next decade, could detect black-hole binaries at distances
of over \unit[1]{Gpc}~\cite{Fritschel:2003qw}.

Optimal searches for gravitational waves use matched filtering, which
requires accurate knowledge of the waveform~\cite{thorne.k:1987}.
Previous searches in LIGO data have used post-Newtonian and
phenomenological templates to search for the coalescence of black-hole
binaries~\cite{Abbott:2005pf,Abbott:2007xi,Abbott:2008}. Over the last
several years numerical relativity has made remarkable breakthroughs
in simulating the late inspiral, merger and ringdown of black-hole
binaries. The computational cost of these simulations is high,
however, making it impractical to use them directly as template
waveforms for use in a matched-filter search. It has been shown that
there is good agreement between the waveforms generated by numerical
relativity with analytic post-Newtonian waveforms to within just a few
orbits of merger~\cite{Buonanno-Cook-Pretorius:2007, Baker2006d,
  Pan2007, Buonanno2007, Hannam2007, Boyle2007, Gopakumar:2007vh,
  Hannam2007c, Boyle2008a, Mroue2008, Hinder2008b}.

This paper uses the high-accuracy Caltech--Cornell
numerical-relativity waveforms to suggest improvements to the analytic
waveforms currently used in gravitational-wave searches by LIGO and
Virgo.  A similar study has been performed by Pan~\etal using
numerical data from Pretorius and the Goddard groups~\cite{Pan2007}.
Our main results are in agreement with their conclusion that a simple
extension of the existing stationary-phase approximation to the
adiabatic post-Newtonian waveforms (called \textit{TaylorF2} in
Ref.~\cite{Damour2001}) yields high overlaps with numerical waveforms.

In Sec.~\ref{sec:Searches}, we review the current techniques used for
searching for gravitational waves in gravitational-wave detector data.
We discuss the construction of the waveform---a pN--NR hybrid---in
Sec.~\ref{sec:PNNRHybridWaveform}.  In Sec.~\ref{sec:Efficiency} we
employ the hybrid waveform in a comparison of the detection efficiency
of gravitational-wave templates that may be used in upcoming searches
of LIGO and Virgo data.  Finally, in Sec.~\ref{sec:Recommendations},
we discuss improvements that may be made to the current data-analysis
techniques to optimize overlaps.

Throughout this paper, we use only the $(l,m)=(2,2)$ component of the
waveform $\Psi_{4}^{2,2}$ (as defined, e.g., in~\cite{Boyle2008a}).
For convenience, we drop the superscript.  Whenever possible, we use
dimensionless quantities, like $r\,M\,\lvert \Psi_{4} \rvert$, where
$r$ is the areal radius of the observation sphere, and $M$ is the
total apparent-horizon mass of the holes in the initial data.
However, for any calculation involving the LIGO noise curve, we have a
physical scale, and thus use standard mks units.


\section{Searches for gravitational waves from black-hole binaries}
\label{sec:Searches}

\subsection{Matched filtering}
\label{sec:MatchedFiltering}

Current searches for gravitational waves from binary black-hole
coalescence use matched filtering to search for a waveform buried in
noise.  The matched filter is the optimal filter for detecting a
signal in stationary Gaussian noise.  Suppose that $n(t)$ is a
stationary Gaussian noise process with one-sided power spectral
density $S_n(f)$ given by $\langle \tilde{n}(f) \tilde{n}^\ast(f')
\rangle=\frac{1}{2} S_n(|f|)\delta(f-f')$.  For long integration
times, the data stream $s(t)$ output by the detector will always be
dominated by the noise.  Thus, we can simply approximate $n \approx s$
to calculate $S_{n}(f)$.

Using this power spectral density (PSD), we can define the inner
product between two real-valued signals---the data stream $s$ and the
filter template $h$---by
\begin{eqnarray}
  \label{eq:InnerProduct}
  \InnerProduct{s|h} &\equiv 2\, \Re \int_{-\infty}^{\infty}\,
  \frac{\tilde{s}(f)\, \tilde{h}^{\ast}(f)}{S_{n}(\lvert f
    \rvert)}\, d f \\ &= 4\, \Re \int_{0}^{\infty}\,
  \frac{\tilde{s}(f)\, \tilde{h}^{\ast}(f)}{S_{n}(f)}\, d f\ .
\end{eqnarray}
Then, given data $s$ which may contain either noise $n$ or noise and a
gravitational wave signal $h$,
\begin{equation}
  s = \left\{\begin{array}{l}
      n  \\
      n+h
    \end{array} \right.\ ,
\end{equation}
the matched-filter signal-to-noise ratio (SNR) is defined as
\begin{equation}
  \label{eq:InnerProductSNR}
  \rho = \frac{1}{\sqrt{\InnerProduct{h|h}}} \InnerProduct{s|h}\ .
\end{equation}

The SNR can then be used to construct a detection statistic (directly
or in combination with other statistics).  It is therefore important
to ensure that the templates used in searches accurately model the
expected waveforms to avoid reduction in the value of $\rho$. The
\emph{overlap} between two templates $h$ and $h'$ is defined as
\begin{equation}
  \label{eq:OverlapDefinition}
  \Overlap{h|h'} \equiv \frac{\InnerProduct{h|h'}}{
    \sqrt{\InnerProduct{h|h} \InnerProduct{h'|h'}}}\ .
\end{equation}
The overlap encodes the fractional loss in SNR that results from using
the template $h'$ rather than the true waveform $h$.  In a search that
uses $\rho$ as a detection statistic this corresponds to the
fractional loss in distance to which the search is sensitive.

The filter template includes arbitrary time and phase offsets, encoded
by the arrival time and phase, $\ta$ and $\phia$.  Under a change of
these quantities, the Fourier transform behaves as
\begin{equation}
  \label{eq:EffectOfTimeAndPhaseOffset}
  \tilde{h}(f) \to \tilde{h}(f)\, \e^{-2\pi i f \ta - i \phia}\ .
\end{equation}
We maximize over these two variables by calculating the inner product
as

\begin{eqnarray}
  \max_{\ta, \phia}\, \InnerProduct{s|h}
  &= \max_{\ta, \phia}\, 4\, \Re \int_{0}^{\infty}\,
  \frac{\tilde{s}(f)\, \tilde{h}^{\ast}(f)}{S_{n}(f)}\, \e^{2\pi i
    f\ta + i \phia}\, d f
  \\
  & = 4 \max_{\ta}\, \left\lvert \int_{0}^{\infty}\,
    \frac{\tilde{s}(f)\, \tilde{h}^{\ast}(f)}{S_{n}(f)}\, \e^{2\pi i
      f\ta}\, d f \right\rvert\ .
\end{eqnarray}
Note that this integral is just the (inverse) Fourier transform of the
quantity $\tilde{s}(f)\, \tilde{h}^{\ast}(f) / S_{n}(f)$ evaluated at
$\ta$.  Thus finding the maximum over $\ta$ involves taking the
Fourier transform and selecting the largest element of the finite set
that results from discretization.

In this paper we are concerned with overlaps between pN waveforms and
NR signals; to weight the inner product we use the following PSDs for
Initial and Advanced LIGO: for Initial LIGO we use an analytic
approximation to the LIGO design PSD given by
\begin{eqnarray}
  S_n(f) &= &3.136 \times 10^{-4} \bigg[
  \left(\frac{ 4.49 f}{150.0}\right)^{-56.0} \nonumber \\
  &+ & 0.16 \left(\frac{f}{150}\right)^{-4.52}
  + \left(\frac{f}{150.0}\right)^2 + 0.52
  \bigg]
\end{eqnarray}
All integrals start from 40 Hz.  As shown in
Fig.~\ref{fig:StildesAndInitialPSD}, at this frequency the noise is an
order of magnitude higher than its lowest value, and below this
frequency it rises rapidly as $\sim f^{-56}$.  The region below 40 Hz
therefore contributes very little signal power to the
SNR~\cite{Abbott:2007xi}.  The PSD for Enhanced LIGO, which will
begin operation in mid 2009, has a similar shape to that for Initial
LIGO although it has a factor of $\sim 2$ increase in strain
sensitivity.  Our results using the Initial-LIGO PSD are therefore
valid for Enhanced LIGO, as the sensitivity factor cancels in
Eq.~\eqref{eq:OverlapDefinition}; overlaps depend on the \emph{shape}
of the PSD.

For Advanced LIGO we use the GWINC program~\cite{AdvancedLIGONoise} to
generate the PSD.  Bench reports the PSD in increments of 0.0124 Hz.
When calculating discrete integrals against signals sampled at other
frequencies we obtain values for the PSD by linearly interpolating
between the values provided by Bench.  We start integrals at 10 Hz as
that is the point where the noise has increased by two orders of
magnitude above its minimum, as also shown in
Fig.~\ref{fig:StildesAndInitialPSD}.

\subsection{Post-{N}ewtonian template}
\label{sec:PNTemplates}

Searches for gravitational waves in LIGO and Virgo use a
post-Newtonian waveform known
as\textit{TaylorF2}~\cite{Abbott:2008}. This is a frequency-domain
waveform obtained via the stationary-phase
approximation~\cite{CutlerFlanagan1994}, which assumes that the
frequency-domain amplitude is simply proportional to $f^{-7/6}$ (the
lowest-order behavior), while its phasing is given by the phase of the
time-domain waveform, as a function of frequency.  For a binary
consisting of masses $m_{1}$ and $m_{2}$, located at an ``effective''
distance $\deff$, we have
\begin{equation}
  \tilde{h}(f; M, \eta, f_{c}) =
  \Theta(f_c - f)\, \left(\frac{\unit[1]{Mpc}}{\deff}\right)\,
  {\mathcal{A}}_{\unit[1]{Mpc}}(M,\eta)\, f^{-7/6}\,
  \e^{i\varphi(f;M,\eta)}\ ,
  \label{e:waveform}
\end{equation}
where
\begin{eqnarray} {\mathcal{A}}_{\unit[1]{Mpc}}(M,\eta) &= &
  \left(\frac{5\pi}{24}\right)^{1/2}
  \left(\frac{\G \MSun/\c^2}{\unit[1]{Mpc}}\right) \times \nonumber \\
  &\quad & \times \left(\frac{\pi \G \MSun}{\c^3}\right)^{-1/6}
  \left(\frac{\eta}{\MSun}\right)^{1/2}
  \left(\frac{M}{\MSun}\right)^{1/3}\ ,
\end{eqnarray}
and the phasing $\varphi$ of the frequency-domain waveform is given to
3.5pN accuracy by the formula~\cite{Blanchet.2002,Blanchet.2004}
\begin{eqnarray}
  \varphi(f;M,\eta) &=& 2\pi ft_0-2\phi_0-\pi/4 \nonumber \\
  &&+ \frac{3}{128\eta}\Bigg[v^{-5}
  +\left(\frac{3715}{756}+\frac{55}{9}\eta\right)v^{-3}-16\pi v^{-2} \nonumber \\
  &&+ \left(\frac{15\,293\,365}{508\,032}+\frac{27\,145}{504}\eta
    +\frac{3085}{72}\eta^2\right)v^{-1} \nonumber \\
  &&+ \pi \left[ \frac{38\,645}{756} -
    \frac{65}{9}\eta \right] \left[1 + 3
    \ln\left(\frac{v}{v_0}\right) \right] \nonumber \\
  &&+ \left[\frac{11\,583\,231\,236\,531}{4\,694\,215\,680}
    - \frac{640}{3} \pi^2 - \frac{6\,848}{21} \gamma \right] v \nonumber \\
  &&+ \bigg[\left(-\frac{15\,335\,597\,827}{3\,048\,192} +
    \frac{2\,255}{12} \pi^2 - \frac{47\,324.0}{63.0} -
    \frac{794\,8}{9}\right) \eta \nonumber \\
  & & \qquad + \frac{76\,055}{1\,728} \eta^2
  - \frac{127\,825}{1\,296} \eta^3 \bigg] v \nonumber \\ 
  &&+ \pi
  \left[ \frac{77\,096\,675}{254\,016} + \frac{378\,515}{1\,512}
    \eta - \frac{74\,045}{756} \eta^2 \right] v^2\Bigg]\ .
\end{eqnarray}
where $v = \left(\frac{\G M}{\c^3}\pi f\right)^{1/3}$, $M = m_{1} +
m_{2}$ and $\eta =m_1 m_2/(m_1+m_2)^2$.

The overall frequency scale is set by the total mass $M$, as can be
seen by observing that each occurrence of $f$ is accompanied by a
factor of $M$.\footnote{The term $2\pi f t_{0}$ might be rewritten as
  $2\pi Mf \times t_{0}/M$.  This term accounts for a time offset
  altering the phase of the Fourier transform.}  %
Thus, going to a higher-mass system shifts the waveform to lower
frequencies.  On the other hand, to first order, the timescale for the
rate of change of the frequency is given by the \textit{chirp mass}:
\begin{equation}
  \label{eq:ChirpMass}
  \mathcal{M} \equiv \left( \frac{m_{1}^{3}\, m_{2}^{3}} {m_{1} +
      m_{2}} \right)^{1/5} = M\, \eta^{3/5}\ .
\end{equation}
Clearly, the total mass and chirp mass give us two very different
handles on the behavior of the waveform.  These two handles will be
important when we try to match the template to our waveform in regions
where the post-Newtonian and stationary-phase approximations are poor.
This is typically the case for high-mass systems, which only enter the
detector band late in the inspiral.  In this case, we can still obtain
a high match, at the cost of using templates with the wrong values of
$M$ and $\eta$.

We also note that physical binary systems are restricted to $0 < \eta
\leq 1/4$.  However, for higher values of $\eta$, the formulas shown
above still give plausible waveforms; in fact, in some cases these
templates match the true waveform better than any physical template.
We will explore the implications of allowing unphysical values for
$\eta$ in searches over the templates in
Sec.~\ref{sec:UnrestrictedEta}.

Note the Heaviside function in Eq.~\eqref{e:waveform}.  This contains
a cutoff frequency $f_{c}$ which is used to ensure that the template
does not extend to frequencies much greater than the frequencies
contained in the expected signal.  This is essentially a third
parameter for the template waveform, and will be searched over.  See
Sec.~\ref{sec:EffectOfUpperFreqCutoff} for a discussion of strategies
for optimizing detection by changing this cutoff.

Frequencies which are often used to characterize coalescing binary
black holes are: (i) the frequency at the innermost stable circular
orbit (ISCO), $r=6 M$, around a single Schwarzschild black hole with
the total mass of the binary system; (ii) the frequency at the light
ring, $r=3.0/M$, around a single Schwarzschild black hole with the
total mass of the binary system; (iii) the ringdown frequency of the
final black hole, which depends on both its spin and mass; and (iv) an
``effective ringdown,'' $f_{\mathrm{ERD}} \equiv 1.07\,
f_{\mathrm{Ringdown}}$ defined in~\cite{Pan2007}.

Current searches use 2pN stationary-phase approximation (SPA)
\textit{TaylorF2} templates~\cite{Abbott:2008}.  It has previously
been shown that such waveforms provide acceptable detection templates
for binary neutron stars and sub-solar mass black
holes~\cite{Droz:1999qx}, but not necessarily for higher-mass black
holes.  This is an issue we will investigate below by testing 2pN and
3.5pN templates.

\section{PN--NR hybrid waveform}
\label{sec:PNNRHybridWaveform} %

In order to perform our comparison we need to construct a ``true''
black-hole binary waveform, which we might expect to observe with
detectors.  A numerical simulation will provide the data for the
crucial nonlinear merger phase.  We carefully extract the data and
extrapolate it to large radius, and investigate the effects of
numerical error on the final result.  Because this waveform is very
computationally expensive to produce, it covers only about 32 cycles,
which is not sufficient for a thorough investigation of the
possibility of detecting it in searches of data from
gravitational-wave detectors.  Thus, we match the numerical waveform
to a post-Newtonian waveform, producing a hybrid which extends for
many thousands of cycles, covering the entire band of interest.

\subsection{Numerical simulation, extraction, and extrapolation}
\label{sec:WaveformExtractionAndExtrapolation}
The numerical simulation is the same as that described in
Refs.~\cite{Boyle2007, Scheel2008}: an equal-mass, non-spinning,
black-hole binary with reduced
eccentricity~\cite{Pfeiffer-Brown-etal:2007}, beginning roughly 16
orbits before merger, continuing through merger and
ringdown~\cite{Scheel2008}.  It is performed with the Caltech--Cornell
pseudospectral code, using boundary conditions designed to prevent
constraint violations and gravitational radiation from entering the
domain~\cite{Holst2004, Lindblom2006}.

Data is extracted from the simulation in the form of the
Newman--Penrose scalar
\begin{equation}
  \Psi_{4} = -C_{\alpha \beta \gamma \delta} l^{\alpha}
  \bar{m}^{\beta} l^{\gamma} \bar{m}^{\delta}\ ,
\end{equation}
where $l^{\alpha}$ and the complex vector $\bar{m}^{\beta}$ are
constructed with reference to the coordinate basis.  Along the
positive $z$ axis, we have
\begin{eqnarray}
  l^{\alpha} &= & \frac{1}{\sqrt{2}}\, \left( t^{\alpha} - z^{\alpha}
  \right)\ , \\
  \bar{m}^{\beta} &= & \frac{1}{\sqrt{2}}\, \left(
    \frac{\partial}{\partial x} - i\, \frac{\partial}{\partial y}
  \right)^{\beta}\ .
\end{eqnarray}

Here, $t^{\alpha}$ is the timelike unit normal to the spatial
hypersurface, and $z^{\alpha}$ is the unit vector in the positive $z$
direction.  The vectors $\partial/\partial x$ and $\partial/\partial
y$ are the standard coordinate vectors, which are not normalized.
$\Psi_{4}$ is extracted as a function of time, at various radii along
the positive $z$ axis.  This is then extrapolated to large radii, as
described in Ref.~\cite{Boyle2007}, and in greater detail in
Ref.~\cite{Boyle2008}.


The measured (instantaneous) frequency at the beginning of the
simulation is
\begin{equation}
  \label{eq:MeasuredStartingFreq}
  f_{\mathrm{initial}} = \unit[(1.08 \pm 0.01) \times 10^{3}]{Hz}\,
  \frac{\MSun}{M}\ .
\end{equation}
The measured ringdown frequency is
\begin{equation}
  \label{eq:MeasuredRingdownFreq}
  f_{\mathrm{ringdown}} = \unit[(1.78 \pm 0.02) \times 10^{4}]{Hz}\,
  \frac{\MSun}{M}\ .
\end{equation}
The measured \emph{Christodoulou} mass and spin of the final black
hole are
\begin{eqnarray}
  \label{eq:MeasuredFinalMass}
  M_{\chi\mathrm{, final}} &= &(0.95162 \pm 0.00002)\, M_{\chi\mathrm{,
      initial}}\ , \\
  \label{eq:MeasuredFinalSpin}
  S_{\mathrm{final}} &= &(0.68646 \pm 0.00004)\, M^{2}_{\chi\mathrm{, final}}\ .
\end{eqnarray}
Using this value for the spin, a quasi-analytic formula due to
Echeverria~\cite{Echeverria1989} predicts a value of
$\unit[1.77\times10^{4}]{Hz}\, \frac{\MSun}{M}$, for the ringdown
frequency, in close agreement with the measured frequency.

\subsection{Accuracy of the numerical simulation}
\label{sec:Accuracy}

\begin{figure}
  \begin{center}
    \includegraphics[width=0.55\linewidth]{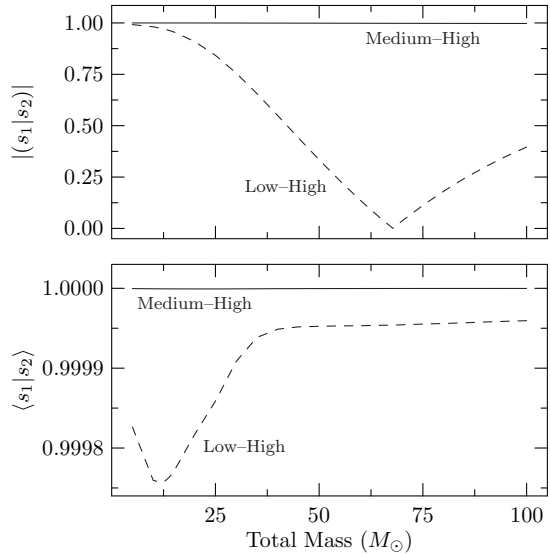}
  \end{center}
  \caption{Convergence testing for numerical waveforms from a
    data-analysis perspective, using the match between waveforms
    computed at different numerical resolutions.  The waveforms are
    scaled to various masses, and the Initial-LIGO noise curve is used
    in the calculation of the match.  The upper panel shows the
    overlap without maximization over arrival time and phase; the
    lower panel shows the overlap after maximization.  In each panel,
    the lower (dashed) line compares the lowest- and
    highest-resolution simulations, while the upper (solid) line
    compares the medium- and highest-resolution simulations.  Note
    that this plot uses only numerical data, with no post-Newtonian
    contribution.}
  \label{f:accuracy}
\end{figure}

The numerical waveform will be the standard against which we will
judge the \textit{TaylorF2} waveforms used in LIGO data analysis.  To
understand how precisely we should trust our final results, we need to
understand the accuracy of the waveform itself.  The most obvious
measure of the error in this fiducial waveform is its convergence with
increasing numerical resolution.  Fig.~\ref{f:accuracy} shows the
overlap (Eq.~\eqref{eq:OverlapDefinition}) between waveforms computed
at different resolutions.  The data used here are the extrapolated
$\Psi_4$ waveforms, integrated in time twice.

Because of the short extent of the numerical waveforms, we need to be
careful when using their Fourier transforms.  The signal can be
corrupted easily by the non-periodicity of the waveforms, and the
discontinuous jumps that result.  For Fig.~\ref{f:accuracy} we
mitigate this problem by increasing the sampling frequency of the
input data, and restricting the Fourier transform to frequencies
corresponding to instantaneous frequencies contained in the data.  The
input data can easily be upsampled in the time domain by interpolating
the phase and amplitude of the complex data to a finer time grid.  We
then perform the transform, and explicitly set the data to zero at
frequencies below $f_{\mathrm{initial}}$ and above
$f_{\mathrm{ringdown}}$, as given in
Eqs.~\ref{eq:MeasuredStartingFreq} and~\ref{eq:MeasuredRingdownFreq}.
While the results do depend on whether or not we impose these cutoffs,
they do not depend sensitively on the actual cutoff frequencies.

The overlap between the lowest- and highest-resolution simulations
(dashed lines) actually passes through zero, as shown in the upper
panel.  Presumably, this is because of loss of phase accuracy over the
course of the simulation.  All three simulations begin with the same
initial data, so the waveforms are most similar at the beginning.
Masses for which this is the most important segment (the lowest
masses) will naturally have the highest overlap between resolutions.
As the simulation progresses, numerical error accumulates---notably in
the phase---so the overlap decreases with masses for which later
segments dominate the overlap (higher masses).  When the overlap is
optimized over arrival time and phase, we can see that the overlap
becomes much better, as shown in the lower panel, indicating
sufficient accuracy within any frequency band for which phase
coherence is required.  In either case, the medium and
highest resolutions are much more nearly the same.  Without
optimization, their overlap is within a few tenths of a percent of 1;
after optimization, the overlap is within $10^{-6}$ of 1.

In the rest of our analysis we use the highest-resolution waveform.
Because we always optimize over arrival time and phase, the lower
panel of Fig.~\ref{f:accuracy} is the most relevant, and shows that
the waveform has converged to very high accuracy.  The overlaps we
quote below will only be given to three decimal places at most,
because this is roughly the accuracy of the single-precision numerical
methods used in the rest of the paper.  This accuracy is also
sufficient for searches of gravitational-wave data.  Thus, the
truncation error of the simulated waveform is irrelevant for those
purposes.

Other sources of error include residual eccentricity and spin, the
influence of the outer boundary of the simulation, extrapolation
errors, and coordinate effects, as discussed in Ref.~\cite{Boyle2007}.
The eccentricity had a disproportionately large effect on the error
quoted in that paper because of the matching technique, which is not
used here.  Restricting attention to the other effects of
eccentricity, the uncertainty falls below that due to numerical error.
Similarly, using the techniques of Ref.~\cite{Lovelace2008}, the
initial spins of the black holes have been measured more reliably, and
found to be more than an order of magnitude smaller than previously
determined, allowing us to reduce the estimate for that error to less
than the numerical truncation error.  The various coordinate effects
were all estimated to be of roughly the same magnitude as the
numerical error.

With the numerical error being many times more accurate than needed
for this analysis, and the other sources of uncertainty being of
roughly the same size, these considerations indicate that the overall
error in our fiducial waveform is substantially less than the
precision needed for this analysis.

\subsection{Hybrid waveform}
\label{sec:HybridWaveform} %
Numerical simulations cannot simulate a very large portion of the
inspiral of a black-hole binary system.  Indeed, the longest such
simulation currently in the literature is the one used here---which
extends over just 32 gravitational wave cycles before merger.
Fortunately, this is the only stage in which simulations are needed.
It has been shown previously~\cite{Boyle2007} that the
\textit{TaylorT4} waveform with 3.5-pN phase and 3.0-pN amplitude
matches the early part of this simulation to very high accuracy.  We
generate a \textit{TaylorT4} waveform of over 8000 gravitational wave
cycles ($t \sim 1.2\times 10^{6}M$, starting at $M f=0.004$), and
transition between the two to create a hybrid.  This long waveform is
sufficient to ensure that---even for the lowest-mass systems we will
consider---the waveform begins well before it enters the frequency
band of interest to LIGO.

We begin with $\Psi_4$ data, which will later be integrated to obtain
$h$.  Following Ref.~\cite{Boyle2008a}, we match the numerical
waveform to the pN waveform by adjusting the time and phase offsets of
the pN waveform to minimize the quantity
\begin{equation}
  \label{eq:MatchingChiSquared}
  \Xi(\Delta t, \Delta \phi) = \int_{t_{1}}^{t_{2}}\, \left[
    \phi_{\NR}(t) - \phi_{\pN}(t - \Delta t) - \Delta \phi
  \right]^{2}\, d t \ .
\end{equation}
Here, we choose $t_{1}=900\,M$ and $t_{2}=1730\,M$, which is closer to
the beginning of the waveform than in the previous paper.  This
particular interval is chosen to begin and end at troughs of the small
oscillations due to the residual eccentricity $e\sim 5\times 10^{-5}$
in our numerical waveform.  Taking a range from trough to trough or
peak to peak---rather than node to node, for example---of the
eccentricity effects minimizes their influence on the matching.  The
eccentricity oscillations can be seen more easily after low-pass
filtering the waveform, though we find filtering to be unnecessary for
this paper.  The junk radiation apparent in the waveform as shown here
has no effect on the resulting match---as we have verified by
filtering, and redoing the match.  Because the final waveform will
incorporate no numerical data before $t_{1}$ and very little
immediately thereafter (as explained below), the junk radiation will
have no effect on any of our results---as we have also explicitly
verified.  In particular, by integrating $\Psi_{4}$ to obtain $h$, we
will effectively smooth the junk radiation.

\begin{figure}
  \begin{center}
    \includegraphics[width=0.55\linewidth]{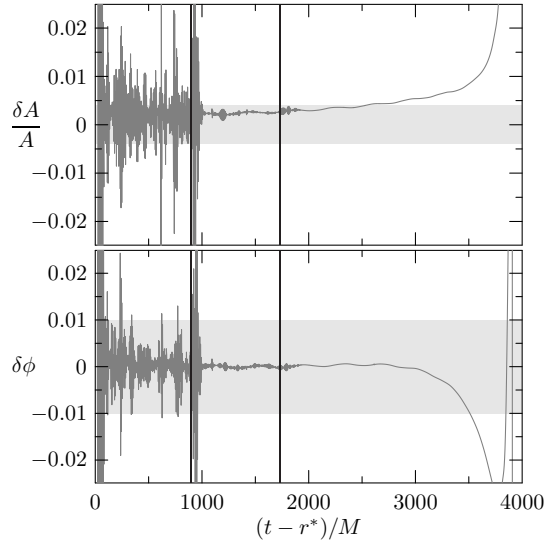}
  \end{center}
  \caption{Amplitude and phase differences between the numerical and
    post-Newtonian waveforms, $\Psi_4$, that are blended to create the
    hybrid waveform.  The vertical lines at $900M$ and $1730M$ denote
    the region over which matching and hybridization occur.  Note that
    the agreement is well within the numerical accuracy of the
    simulation, represented by the horizontal bands, throughout the
    matching region.  Also note that the phase difference is fairly
    flat for a significant period of time after the matching range,
    which indicates that the match is not sensitive to the particular
    range chosen for matching.}
  \label{fig:MatchingPhaseComparison}
\end{figure}%
In Fig.~\ref{fig:MatchingPhaseComparison} we compare the phase of the
numerical and pN waveforms.  The quantities plotted are
\begin{eqnarray}
  \delta \phi & \equiv & \phi_{\pN} - \phi_{\NR}\ , \\
  \frac{\delta A}{A} & \equiv & \frac{A_{\pN} - A_{\NR}}{A_{\NR}}\ , 
\end{eqnarray}
shown over the interval on which both data sets exist.  The vertical
bars denote the matching region.  Note that the phase difference is
well within the accuracy of the simulation (about 0.01 radians,
represented by the horizontal band) over a range extending later than
the matching region.  Also, the difference between the two is fairly
flat, which implies that the match is not very sensitive to the region
chosen for matching.  Because of this, we expect that the phase
coherence between the early pN data and the late NR data will be
physically accurate to high precision.

The hybrid waveform is then constructed by blending the two matched
waveforms together according to
\begin{eqnarray}
  \label{eq:HybridWaveform}
  A_{\hyb}(t) &= & \tau(t)\, A_{\NR} + \left[ 1 - \tau(t) \right]\,
  A_{\pN}(t)\ , \\
  \phi_{\hyb}(t) &= & \tau(t)\, \phi_{\NR} + \left[ 1 - \tau(t)
  \right]\, \phi_{\pN}(t)\ .
\end{eqnarray}
The blending function $\tau$ is defined by

\begin{equation}
  \label{eq:BlendingFunction}
  \tau(t) = \left\{\begin{array}{ll}
      0 & \mathrm{if}\quad t<t_{1}  \\
      \frac{t-t_{1}}{t_{2}-t_{1}} & \mathrm{if}\quad t_{1} \leq t < t_{2} \\
      1 & \mathrm{if}\quad t_{2} \leq t
    \end{array} \right.
\end{equation}

\begin{figure}
  \begin{center}
    \includegraphics[width=0.55\linewidth]{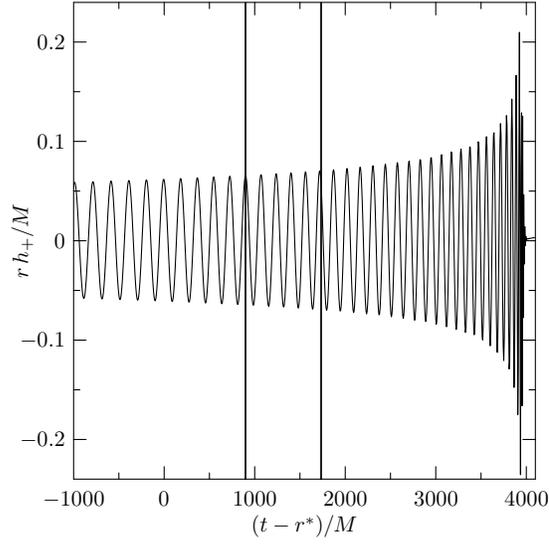}
  \end{center}
  \caption{The last $t=5000\MSun$ of the hybrid waveform used in this
    analysis: the $h_{+}$ waveform seen by an observer on the positive
    $z$ axis.  The vertical lines denote the matching and
    hybridization region.  The $0$ on the time axis corresponds to the
    beginning of data from the numerical simulation.}
  \label{fig:WaveformSnapshot}
\end{figure}

The values of $t_{1}$ and $t_{2}$ are the same as those used for the
matching.  The amplitude discrepancy between the pN waveform and the
NR waveform over this interval is within numerical
uncertainty---roughly $0.4\%$.  As with the matching technique
(Eq.~(\ref{eq:MatchingChiSquared})), this method is similar to that of
Ref.~\cite{Ajith-Babak-Chen-etal:2007b}, but distinct, in that we
blend the phase and amplitude, rather than the real and imaginary
parts.  This leads to a smoothly blended waveform, shown in
Fig~\ref{fig:WaveformSnapshot}.

Up to this point, the waveform has been $\Psi_{4}$ data.  With the
long waveform in hand, we numerically integrate twice to obtain $h$,
and set the four integration constants so that the final waveform has
zero average and first moment~\cite{Pfeiffer-Brown-etal:2007}.
Because of the very long duration of the waveforms, this gives a
reasonable result, which is only incorrect at very low
frequencies---lower than any frequency of interest to us.  We have
also checked that our results do not change when we effectively
integrate in the frequency domain by taking
\begin{equation}
  \label{eq:PsiFourIntegration}
  \tilde{h} = -\frac{\tilde{\Psi}_{4}}{4\,\pi\, f^{2}}\ ,
\end{equation}
which is the frequency-domain analog of the equation $\Psi_{4} =
\ddot{h}$.

\section{Detection efficiency of gravitational-wave templates}
\label{sec:Efficiency} %

We now compare the signal described in the previous section to
restricted, stationary phase \textit{TaylorF2} post-Newtonian
templates with terms up to order 2.0, order 3.5, and a ``pseudo-4.0
pN-order'' term recommended in Ref.~\cite{Pan2007}.  Overlaps are
calculated using the techniques of Sec.~\ref{sec:MatchedFiltering},
with the signal $s$ being the hybrid waveform described in
Sec.~\ref{sec:PNNRHybridWaveform} scaled to a range of masses.  We
consider both the Initial- and Advanced-LIGO noise curves.

\begin{figure}
  \begin{center}
    \includegraphics[width=0.55\linewidth]{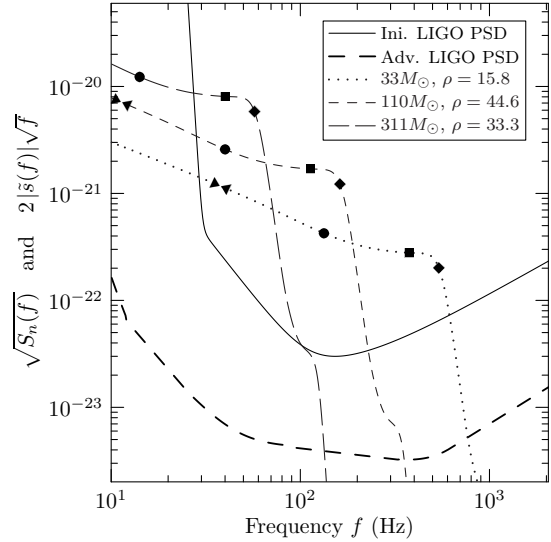}
  \end{center}
  \caption{Hybrid Caltech--Cornell waveform scaled to various total
    masses, with sources optimally oriented and placed at
    \unit[100]{Mpc}, shown against the Initial- and Advanced-LIGO
    noise curves.  Markers are placed along the lines at frequencies
    corresponding to various instantaneous frequencies of the
    waveforms.  The triangles represent the beginning and end of the
    blending region; the circle represents the ISCO frequency; the
    square the light-ring; and the diamond the measured ringdown
    frequency.  See the text for discussion of the normalization.  The
    values given for $\rho$ use the Initial-LIGO noise curve, with
    sources at a distance of 100\,Mpc.}
  \label{fig:StildesAndInitialPSD}
\end{figure}

Plots of the hybrid waveforms in comparison to the Initial-LIGO noise
curve are shown in Fig.~\ref{fig:StildesAndInitialPSD}.
The masses are chosen so that various frequencies of interest (the
final stitching frequency, the ISCO, and the ringdown) occur at the
``seismic wall'' for Initial LIGO: \unit[40]{Hz}.  The waveforms
$\tilde{s}$ are scaled to depict the detectability of the signal,
typically quantified by the SNR introduced in
~\eqref{eq:InnerProductSNR}, which may be written as
\begin{equation}
  \label{eq:SNR}
  \rho^{2} \equiv \int_0^\infty \frac{4\, \tilde{s}(f)\,
    \tilde{s}^\ast(f)} {S_n(f)}\, d f = \int_0^\infty
  \frac{\left\lvert 2\, \tilde{s}(f)\, \sqrt{f} \right\rvert^{2}}
  {S_n(f)}\, d\ln{f}\ .
\end{equation}
In the final expression, the numerator and denominator have the same
units, and are directly comparable.  Because the square root of the
denominator is familiar, we plot that along with the square root of
the numerator.  Plotting these two quantities together gives a
graphical impression of the detectability of the waveform, and the
relative importance of each part of the waveform, by its height above
the noise curve.  In Ref.~\cite{BradyCreighton2002}, Brady and
Creighton define a slightly different quantity, the characteristic
strain $h_{\mathrm{char}} \equiv f\, \lvert \tilde{s}(f) \rvert\ .$
The relative factor of $\sqrt{f}$ they use is present so that they can
plot $h_{\mathrm{char}}$ against $\sqrt{f\, S_{n}(f)}$.  Cutler and
Thorne~\cite{Cutler2002} define still another quantity, the signal
strength $\tilde{h}_{s}(f)$, which is related to the Fourier transform
by $\tilde{h}(f) = \sqrt{5}\, \frac{T}{N}\, \tilde{h}(s)\ .$ The
factor of $\sqrt{5}$ comes from averaging over the orientation of the
binary, which we do not do.  $T/N$ is the ratio of the threshold to
the rms noise at the endpoint of signal processing.

For each template family we initially optimize over signal mass $M$,
symmetric mass ratio $\eta = m_1 m_2 / (m_1 + m_2)^2$, and upper
cutoff frequency $f_c$.  The optimization is performed using a
Nelder--Mead (``amoeba'') algorithm~\cite{numrec_cpp}.  The amoeba
starts with a simplex in the parameter space, and proceeds through a
series of steps, each of which will improve the value of the function
at at least one vertex.  The algorithm terminates when all vertices
have converged to the same point to within a specified tolerance.
This process is deterministic, and amounts to an enhanced
steepest-ascent algorithm.  It is therefore only guaranteed to find a
local maximum, and indeed we find that an amoeba instance started at a
random point in the parameter space is most likely to converge to a
point that does not give the highest possible overlap.  We interpret
this as being due to a large region in parameter space containing a
local maximum and a relatively smaller region containing the global
maximum.  We therefore supplement the basic amoeba by running 300
instances with random starting values, and taking the best match
obtained over all instances.  In repeated runs the same optimal
parameters were found by at least some of the amoebas, which supports
the claim that this is the true maximum.

\begin{figure}
  \includegraphics[width=0.5\linewidth]{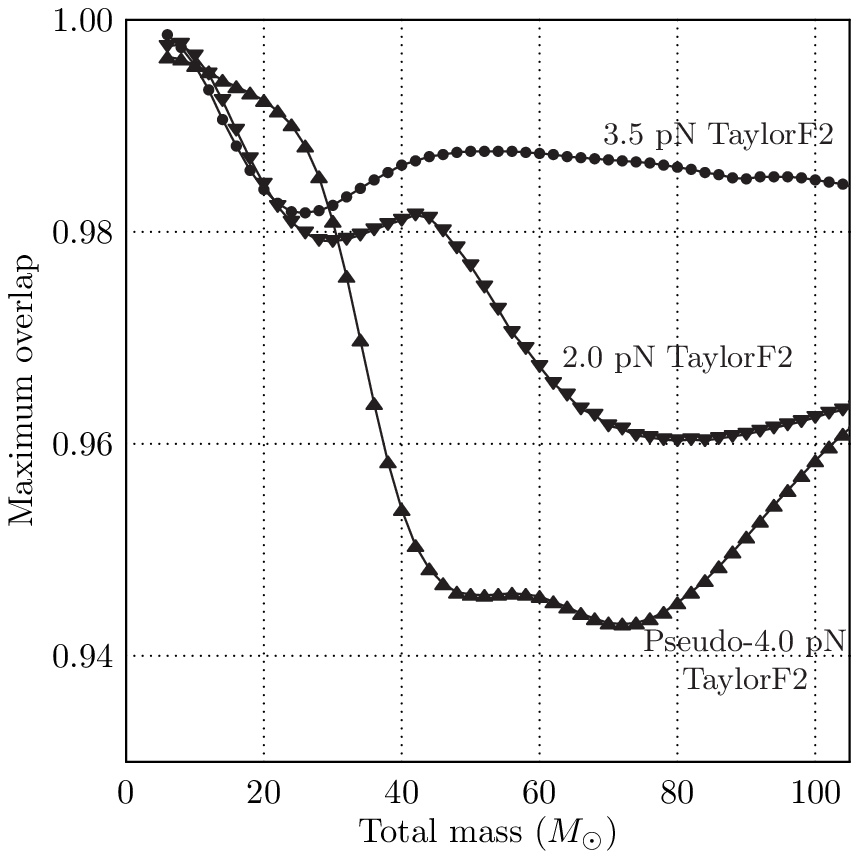}
  \includegraphics[width=0.5\linewidth]{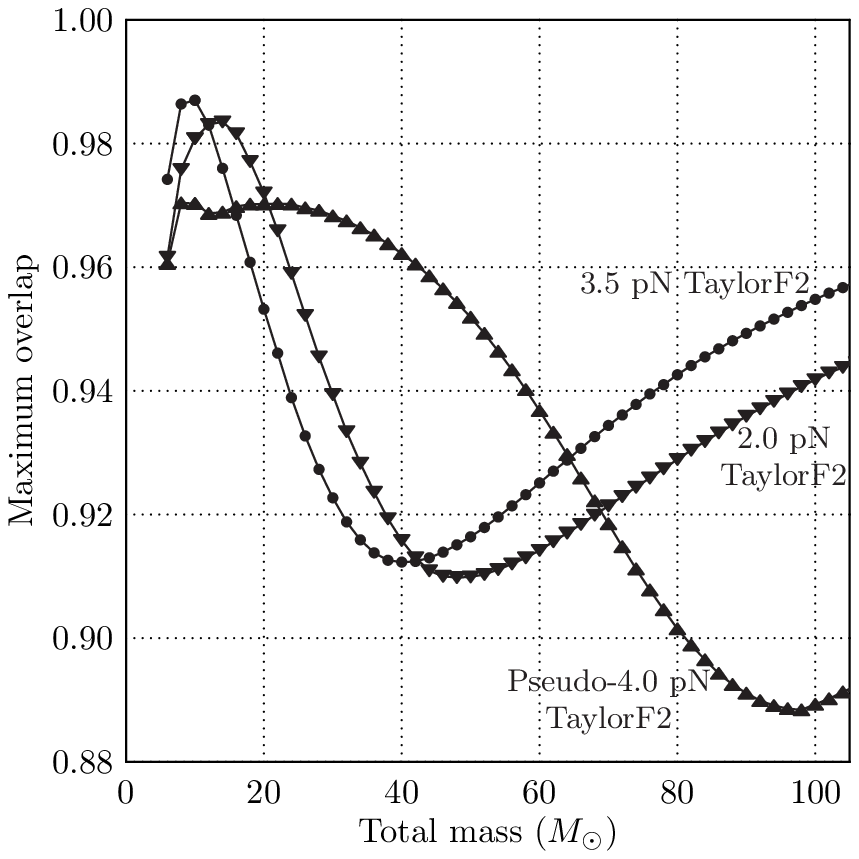}
  \caption{Left: Overlaps between Caltech--Cornell hybrid waveforms,
    scaled to various masses, and restricted stationary-phase pN
    waveforms for Initial-LIGO PSD. Optimization is over $M$ and
    $\eta$, which the cutoff frequency $f_{c}$ is prescribed by the
    weighted average described below.  The mass ratio $\eta$ is
    allowed to range over unphysical values.  The best-fit values
    found for the pseudo-4.0 pN templates are always physical in this
    case.  See Sec.~\ref{sec:UnrestrictedEta}.  Right: The same, for
    the Advanced-LIGO PSD}
  \label{fig:ThreeParamOverlapSummaries}
\end{figure}

The results of optimizing over all of $M, \eta$ and $f_c$ for selected
masses for Initial LIGO are given in
Table~\ref{tab:ThreeParamOverlapDetailInitial} and summarized in
Fig.~\ref{fig:ThreeParamOverlapSummaries}.  For Initial LIGO, in the
range covered by the current Compact Binary Coalescence (CBC) low-mass
search $(M < 35 \MSun)$~\cite{Abbott:2008}, the pseudo-4.0 pN
\textit{TaylorF2} waveforms achieve the highest overlaps, exceeding
those obtained with 3.5 pN waveforms by $\sim 1\%$.  Above $35 \MSun$
the 3.5 pN waveforms produce overlaps as much as 4\% greater than
those obtained with pseudo-4.0 pN waveforms over a range from
$40$--$80 \MSun$. With the Advanced-LIGO noise curve, in the CBC
low-mass range, the 3.5 pN and pseudo-4.0 pN waveforms produce
overlaps within 2\% of each other, with 3.5 pN producing higher
overlaps below 20 $\MSun$ and pseudo-4.0 pN producing higher overlaps
in the range $20$--$35 \MSun$.  Pseudo-4.0 pN continues to give the
highest overlaps up to $60 \MSun$, producing overlaps as much as 4\%
greater than those obtained with 3.5 pN waveforms.  Above $60 \MSun$
3.5 pN waveforms again yield the best overlaps, by as much as 6\%
around 90 $\MSun$.



\begin{table*}
  \begin{center}
    \begin{tabular}{@{}lcccc@{}}
      \hline \hline
      & $(10+10) \MSun$ & $(20+20) \MSun$ & $(30+30) \MSun$ & 
      $(50+50) \MSun$ \\
      \hline
      $\Overlap{s^{\textrm{NR-CC}} |
        h^{\textrm{SPA}_c^{\textrm{ext}}(2.0)}}$ &
      0.99 & 0.98 & 0.97 & 0.96 \\
      $M/\MSun$ &
      $23.27^{+0.13}_{-0.12}$  &
      $25.99^{+0.61}_{-0.56}$  &
      $35.22^{+1.84}_{-1.89}$  &
      $47.52^{+6.87}_{-4.73}$  \\
      $\eta$ &
      $0.199^{+0.0030}_{-0.0030}$  &
      $0.771^{+0.0490}_{-0.0420}$  &
      $1.000_{-0.1390}$  &
      $1.000_{-0.2490}$  \\
      $f_{\mathrm{cut}}$ (Hz) &
      $501.18^{+523.00}_{-153.00}$  &
      $431.35^{+358.00}_{-77.00}$  &
      $296.05^{+53.00}_{-31.00}$  &
      $190.56^{+20.00}_{-14.00}$  \\
      \hline
      $\Overlap{s^{\textrm{NR-CC}} |
        h^{\textrm{SPA}_c^{\textrm{ext}}(3.5)}}$ &
      0.98 & 0.99 & 0.99 & 0.99 \\
      $M/\MSun$ &
      $18.75^{+0.10}_{-0.10}$  &
      $31.88^{+0.77}_{-0.71}$  &
      $47.15^{+4.37}_{-3.27}$  &
      $259.89^{+0.00}_{-194.18}$  \\
      $\eta$ &
      $0.290^{+0.0040}_{-0.0040}$  &
      $0.493^{+0.0530}_{-0.0410}$  &
      $0.756^{+0.2440}_{-0.2290}$  &
      $0.954^{+0.0460}_{-0.2090}$  \\
      $f_{\mathrm{cut}}$ (Hz) &
      $506.50^{+518.00}_{-155.00}$  &
      $448.80^{+576.00}_{-83.00}$  &
      $324.74^{+145.00}_{-42.00}$  &
      $197.17^{+24.00}_{-16.00}$  \\
      \hline
      $\Overlap{s^{\textrm{NR-CC}} |
        h^{\textrm{SPA}_c^{\mathcal{Y}}(4)}}$ &
      0.99 & 0.96 & 0.95 & 0.96 \\
      $M/\MSun$ &
      $23.64^{+0.13}_{-0.12}$  &
      $47.90^{+1.28}_{-1.13}$  &
      $61.81^{+8.68}_{-6.19}$  &
      $89.93^{+20.44}_{-16.60}$  \\
      $\eta$ &
      $0.182^{+0.0030}_{-0.0030}$  &
      $0.181^{+0.0160}_{-0.0140}$  &
      $0.523^{+0.4260}_{-0.1820}$  &
      $0.529^{+0.4720}_{-0.3100}$  \\
      $f_{\mathrm{cut}}$ (Hz) &
      $509.47^{+654.00}_{-145.00}$  &
      $352.44^{+73.00}_{-61.00}$  &
      $309.53^{+72.00}_{-47.00}$  &
      $195.63^{+21.00}_{-15.00}$  \\
      \hline \hline
    \end{tabular}
  \end{center}
  \caption{Maximum overlaps between Caltech--Cornell hybrid waveforms
    and restricted stationary-phase pN templates using the
    Initial-LIGO noise curve.  The first number in each block is the
    overlap; subsequent numbers are the template parameters that
    achieve this overlap.  Parameter values within the specified
    ranges keep the overlap within 1\% of the maximum by varying that
    parameter, while leaving others fixed.  We restrict the search to
    $0 \leq \eta \leq 1.000$, so the upper error bounds when $\eta\sim
    1.000$ may be artificially small.}
  \label{tab:ThreeParamOverlapDetailInitial}
\end{table*}

%
%

\begin{table*}
  \begin{tabular}{@{}lcccc@{}}
    \hline \hline
    & $(10+10) \MSun$ & $(20+20) \MSun$ & $(30+30) \MSun$ & 
    $(50+50) \MSun$ \\
    \hline
    $\Overlap{s^{\textrm{NR-CC}} |
      h^{\textrm{SPA}_c^{\textrm{ext}}(2.0)}}$ &
    0.98 & 0.92 & 0.91 & 0.94 \\
    $M/\MSun$ &
    $25.15^{+0.02}_{-0.02}$ &
    $47.73^{+0.12}_{-0.11}$ &
    $54.39^{+0.51}_{-0.43}$ &
    $60.19^{+1.55}_{-1.29}$ \\
    $\eta$ &
    $0.170^{+0.0010}_{-0.0010}$ &
    $0.188^{+0.0010}_{-0.0010}$ &
    $0.335^{+0.0080}_{-0.0070}$ &
    $0.891^{+0.0660}_{-0.0490}$ \\
    $f_{\mathrm{cut}}$  (Hz) &
    $444.77^{+132.00}_{-115.00}$ &
    $267.64^{+48.00}_{-50.00}$ &
    $262.44^{+34.00}_{-36.00}$ &
    $182.41^{+24.00}_{-18.00}$ \\
    \hline
    $\Overlap{s^{\textrm{NR-CC}} |
      h^{\textrm{SPA}_c^{\textrm{ext}}(3.5)}}$ &
    0.97 & 0.92 & 0.92 & 0.96 \\
    $M/\MSun$ &
    $20.27^{+0.02}_{-0.02}$ &
    $38.11^{+0.11}_{-0.09}$ &
    $50.09^{+0.49}_{-0.42}$ &
    $78.10^{+1.89}_{-1.50}$ \\
    $\eta$ &
    $0.245^{+0.0010}_{-0.0010}$ &
    $0.277^{+0.0020}_{-0.0020}$ &
    $0.386^{+0.0130}_{-0.0100}$ &
    $0.494^{+0.0760}_{-0.0330}$ \\
    $f_{\mathrm{cut}}$   (Hz) &
    $355.85^{+97.00}_{-88.00}$ &
    $262.83^{+47.00}_{-48.00}$ &
    $281.34^{+41.00}_{-37.00}$ &
    $186.31^{+30.00}_{-19.00}$ \\
    \hline
    $\Overlap{s^{\textrm{NR-CC}} |
      h^{\textrm{SPA}_c^{\mathcal{Y}}(4)}}$ &
    0.97 & 0.96 & 0.94 & 0.90 \\
    $M/\MSun$ &
    $22.24^{+0.02}_{-0.02}$ &
    $46.57^{+0.11}_{-0.11}$ &
    $72.06^{+0.35}_{-0.35}$ &
    $118.50^{+1.99}_{-1.63}$ \\
    $\eta$ &
    $0.208^{+0.0010}_{-0.0010}$ &
    $0.190^{+0.0010}_{-0.0010}$ &
    $0.177^{+0.0020}_{-0.0030}$ &
    $0.186^{+0.0100}_{-0.0070}$ \\
    $f_{\mathrm{cut}}$  (Hz) &
    $473.49^{+551.00}_{-136.00}$ &
    $353.18^{+73.00}_{-69.00}$ &
    $242.43^{+37.00}_{-36.00}$ &
    $152.16^{+19.00}_{-19.00}$ \\
    \hline \hline
  \end{tabular}
  \caption{Maximum overlaps between Caltech--Cornell hybrid waveforms
    and restricted stationary-phase pN templates using the
    Advanced-LIGO noise curve.  The first number in each block is the
    overlap; subsequent numbers are the template parameters that
    achieve this overlap.  Parameter values within the specified
    ranges keep the overlap within 1\% of the maximum by varying that
    parameter, while leaving others fixed.  We restrict the search to
    $0 \leq \eta \leq 1.000$, so the upper error bounds when $\eta\sim
    1.000$ may be artificially small.}
  \label{tab:ThreeParamOverlapDetail}
\end{table*}

A significant feature of
Tables~\ref{tab:ThreeParamOverlapDetailInitial}
and~\ref{tab:ThreeParamOverlapDetail} is the size of the error bars on
the cutoff frequencies.  For $M=20 \MSun$ the cutoff frequency can
vary as much as 128\% above and 28\% below the optimal value while
losing no more than 1\% of overlap. This leads us to consider the
range of possible template parameters which may give high overlaps.
In the next section we consider the reduction in overlap as the
parameters $f_{c}$ and $\eta$ are independently varied from the
optimal value.

\subsection{Effect of upper frequency cutoff}
\label{sec:EffectOfUpperFreqCutoff}

\begin{figure}
  \includegraphics[width=0.50\linewidth]{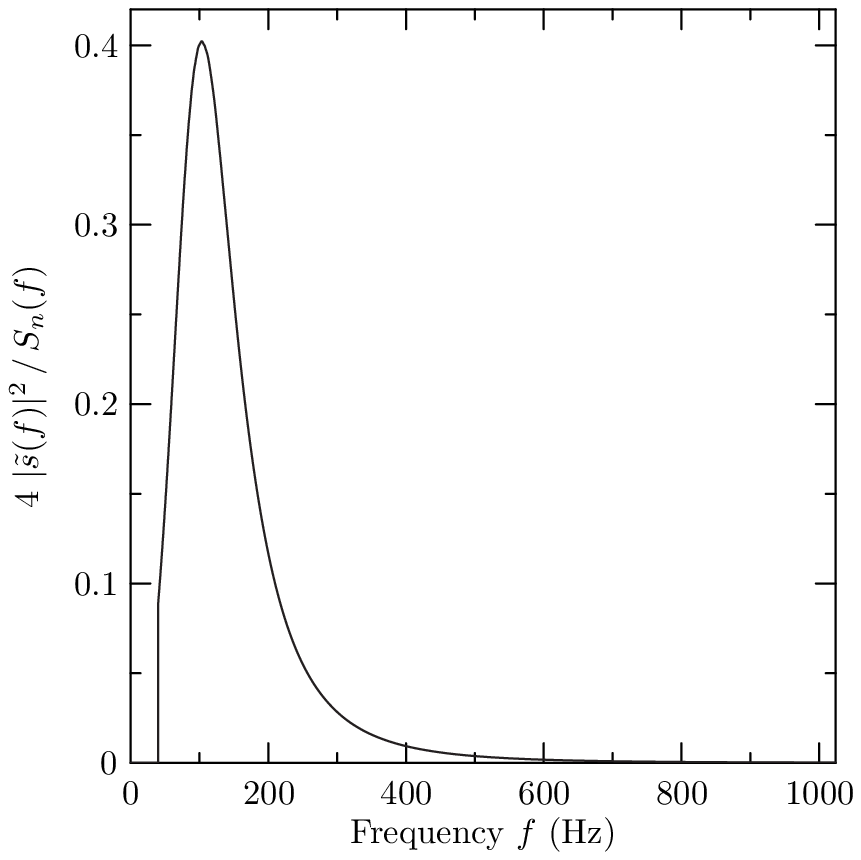}
  \includegraphics[width=0.50\linewidth]{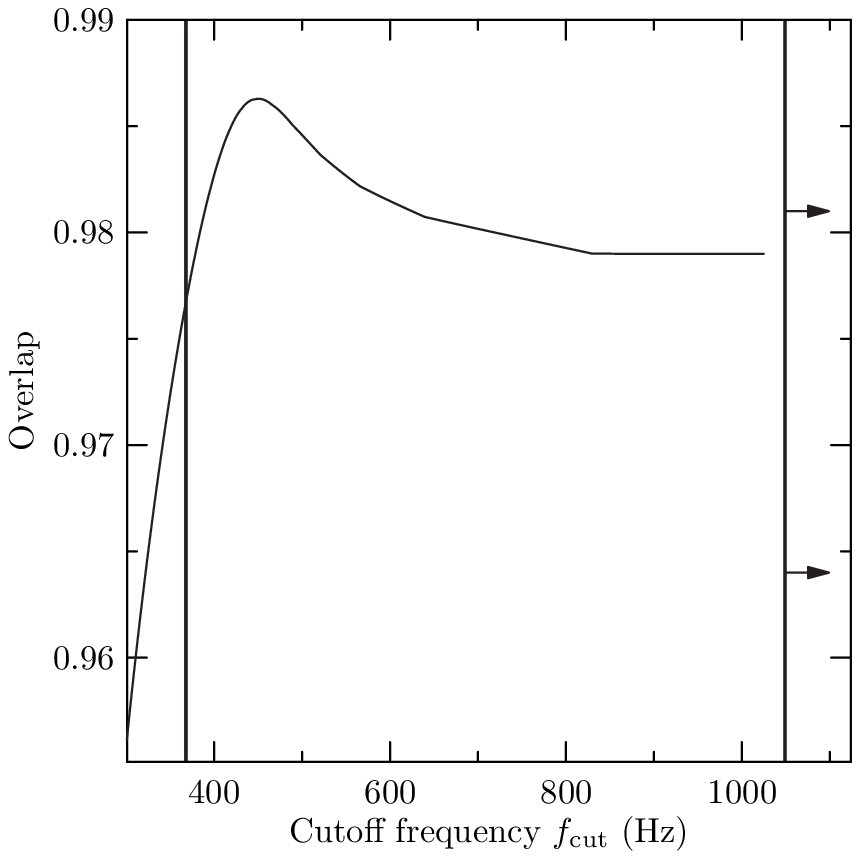}
  \caption{Left: Integrand of Eq.~\eqref{eq:InnerProduct} for a
    \textit{TaylorF2}, 3.5 pN waveform with $M=10$ and $\eta=0.25$, at
    a distance of 100\,Mpc, using the Initial-LIGO noise curve.  Note
    that the shape of this curve does not change as we change $M$ and
    $\eta$; only the vertical scale changes.  Right: Overlap between
    Caltech--Cornell waveform scaled to $M=40\,\MSun$ and restricted
    \textit{TaylorF2}, 3.5 pN waveform using the best-match values for
    $M$ and $\eta$, as a function of the cutoff frequency $f_c$, with
    the Initial-LIGO noise curve.  The vertical bars are meant to
    delineate 1\% loss.  Note that the upper bound extends to higher
    frequencies indefinitely.  }
  \label{fig:FilterIntegrand}
\end{figure}

As shown in Fig.~\ref{fig:StildesAndInitialPSD} the amplitude of the
NR waveforms drops sharply at around the lightring frequency, which
depends on the total mass of the binary.  The \textit{TaylorF2}
waveforms do not model the late inspiral, merger or ringdown and hence
will continue to evolve as $f^{-7/6}$ at all frequencies, increasingly
deviating from the NR waveform.  This suggests that the upper
frequency cutoff of the \textit{TaylorF2} waveform should be chosen to
be below the frequency at which the two diverge.  However, the effect
of the divergence is mitigated by the PSD.  The denominator of the
overlap, Eq.~\eqref{eq:OverlapDefinition}, depends on
$\InnerProduct{s|s}$ which is a constant, and $\InnerProduct{h|h}$
which would increase without limit if not for the PSD.
Fig.~\ref{fig:FilterIntegrand} shows $|\tilde{h}(f)|^2/S_n(f)$---the integrand of
$\InnerProduct{h|h}$---for the Initial-LIGO
noise curve for an example \textit{TaylorF2} waveform for an
equal-mass $10\,\MSun$ binary.  We see that above about 450 Hz there
is very little contribution to the integrand, and so extending the
cutoff frequency above this will not impact the overlap.

The numerator of the overlap, $\InnerProduct{s|h}$, can only increase
as the cutoff frequency is raised, however frequencies above the
lightring where the waveforms have diverged will contribute very
little.  The effect of including higher frequencies on the overlap is
therefore determined by the $\InnerProduct{h|h}$ term in the
denominator.  For systems with ringdown frequencies well above the
peak of the integrand in Fig.~\ref{fig:FilterIntegrand}, this term
will not significantly reduce the overlap.  For example, binaries of
total mass roughly $40\,\MSun$ have ringdown frequencies at roughly
450\,Hz.  Only a small fraction of the SNR comes from higher
frequencies.  Thus, we expect that systems with lower masses should
not suffer great loss in overlap if the cutoff frequency is higher
than ringdown.  However for higher-mass systems the overlap can be
significantly reduced if the upper frequency cutoff is too large.
This is indeed what we find, as shown by a representative example on
the right in Fig.~\ref{fig:FilterIntegrand}.  For this $40\,\MSun$
system, using the Initial-LIGO noise curve, the optimal cutoff
frequency is around 450\,Hz---roughly the ringdown frequency.
Decreasing the cutoff quickly decreases the overlap.  The cutoff may
be increased almost indefinitely, however, with only 0.5\% loss in
overlap.  This, of course, changes when using the Advanced-LIGO noise
curve.  We revisit this issue in Sec.~\ref{sec:Recommendations}.

\subsection{Unrestricted $\eta$}
\label{sec:UnrestrictedEta}

\begin{figure}
  \begin{center}
    \includegraphics[width=0.55\linewidth]{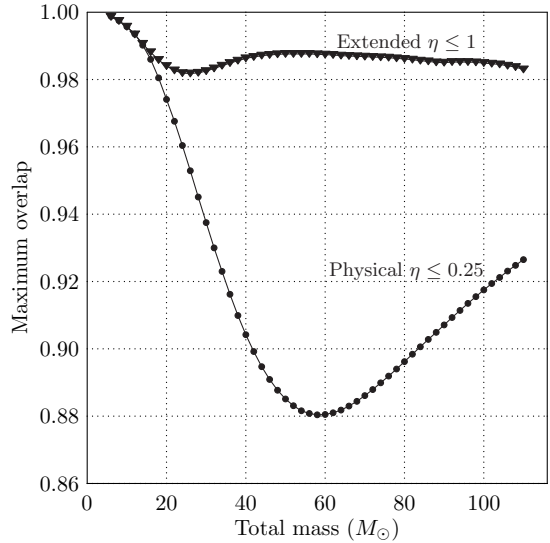}
  \end{center}
  \caption{Maximum overlaps obtained by allowing $\eta$ to range over
    unphysical values, compared to those obtained by restricting the
    range of $\eta$.  These overlaps are generated using 3.5 pN
    TaylorF2 templates, searching over values of the total mass and
    mass ratio.  Extending to unphysical values of $\eta$ improves the
    match by up to 11\%.}
  \label{fig:PhysicalEta}
\end{figure}

The physical symmetric mass ratio is restricted to the range $0 < \eta
\leq 0.25$, values above this imply complex-valued masses.  However
the pN waveforms are well-behaved for $0 < \eta < 1.0$, and as seen
from Tables~\ref{tab:ThreeParamOverlapDetailInitial}
and~\ref{tab:ThreeParamOverlapDetail}, the highest overlaps are often
obtained at unphysical values of $\eta$.  In
Fig.~\ref{fig:PhysicalEta} we show the effect of limiting the
optimization to physical $\eta$.  At high masses, the limitation
reduces the optimal overlap by up to 12\%.  \textit{TaylorF2}
waveforms with $\eta \leq 1/4$ would not be expected to accurately
model the late-inspiral and merger part of the waveform, as
non-Newtonian effects are increasingly significant in this region.  We
find that allowing unphysical $\eta$ broadens the space of waveforms
covered by the \textit{TaylorF2} approximation sufficiently to capture
more of the late-inspiral and merger.

\section{Recommendations for improvements}
\label{sec:Recommendations} %

Based on the analysis of the previous sections we propose a series of
adjustments to searches using \textit{TaylorF2} template waveforms to
enhance the efficiency of those searches.  First, as seen in
Fig.~\ref{fig:ThreeParamOverlapSummaries} for Initial LIGO, adding
terms up to 3.5 pN order produces overlaps as large or larger than the
current 2.0 pN templates over most of the mass range, while the
pseudo-4.0 pN templates recommended in Ref.~\cite{Pan2007} produce
slightly larger overlaps at masses near $20\,\MSun$.  Thus, we
recommend pseudo-4.0 pN templates for the low mass range, $M < 35
\MSun$, and 3.5 pN templates for higher masses.  The improvement due
to 3.5 pN templates over 2.0 pN generally holds for Advanced LIGO as
well.  The 3.5 pN templates produce larger overlaps than 2.0 pN
templates above $50\,\MSun$ without a significant loss (within 1\%) at
lower masses.  However, there is a large region for which the
pseudo-4.0 pN term does significantly better.  When using an
Advanced-LIGO noise curve, we recommend 3.5 pN templates generally,
2.0 pN templates in the range $12$--$21\,\MSun$ and pseudo-4.0 pN
templates for masses in the range $21$--$65\,\MSun$.

As a second improvement, we note from Fig.~\ref{fig:PhysicalEta} that
allowing $\eta$ to range over unphysical values significantly improves
matches with 3.5 pN templates above $30\,\MSun$. In preliminary
studies we have found that extending to $\eta \leq 1$ roughly doubles
the size of the template bank, and the advantages must therefore be
weighed against the increase in false alarm rate.

\begin{figure}
  \includegraphics[width=0.5\linewidth]{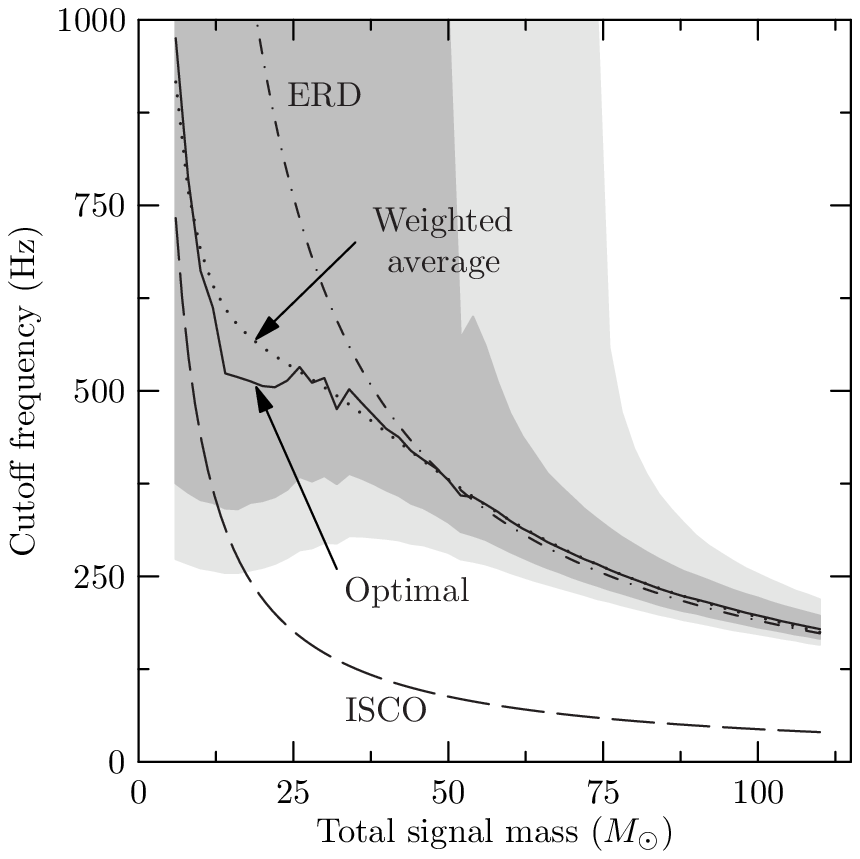}
  \includegraphics[width=0.5\linewidth]{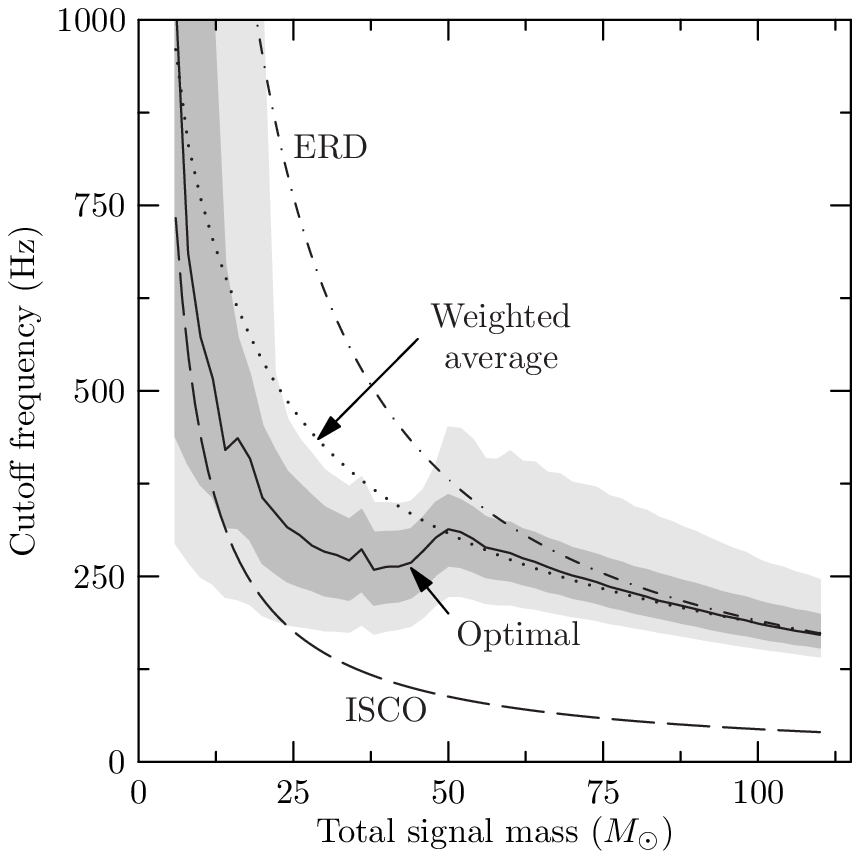}
  \caption{Left: Candidate $f_c$ values for 3.5 pN templates with
    Initial LIGO.  The dark gray band contains cutoff frequencies with
    matches within 1\% of the value at which the best overlap was
    obtained.  The light gray band contains frequencies with matches
    within 3\%.  Right: Candidate $f_c$ values for 3.5 pN templates
    with Advanced LIGO.  The dark gray band contains cutoff
    frequencies with matches within 1\% of the value at which the best
    overlap was obtained.  The light gray band contains frequencies
    with matches within 3\%.  Note that the weighted-average cutoff
    extends past the 1\% error bars for $12 < M/\MSun < 40$.  However,
    in that same region, the 3.5 pN templates do poorly overall, and
    we recommend pseudo-4.0 pN templates.  The optimal cutoff
    frequency for pseudo-4.0 pN templates is much closer to the
    weighted-average cutoff in this mass range.}
  \label{fig:FcRecomendations}
\end{figure}%

Our third recommendation involves the cutoff frequency used for the
template waveform.  Optimization over the cutoff frequency is too
computationally intensive to be done in searches.  Currently, the
cutoff frequency is typically taken to be the Schwarzschild ISCO
frequency.  To examine the effect of this choice we vary $f_c$ while
keeping the mass and $\eta$ at their optimal values, for each of the
signal masses in our range.  The result of one such variation is shown
in Fig.~\ref{fig:FilterIntegrand} (right).
Figs.~\ref{fig:FcRecomendations} shows the variations for all masses,
highlighting the regions within which the overlap drops by less than
1\% (dark gray) and 3\% (light gray) of the optimal value.  This
figure also shows the ISCO and ERD frequencies, neither of which stays
within the 1\% band for both Initial and Advanced LIGO.  In
particular, the ISCO is a poor choice for both Initial and Advanced
LIGO except at very low masses, where the precise value of the cutoff
is almost irrelevant.

The ISCO is often pointed to---somewhat arbitrarily---as a good
estimate of the breakdown of post-Newtonian
approximations~\cite{Blanchet2006}.  So, for instance, if we were to
match a pN template to a physical waveform, beginning at some point in
the distant past, we might expect them to separate quite badly near
the ISCO.  Of course, for realistic black-hole binaries, the
gravitational waves will only enter the LIGO band late in the
inspiral---just before the ISCO for low-mass systems, or after the
ISCO for high-mass systems.  We can see from
Fig.~\ref{fig:StildesAndInitialPSD} that, for masses below about
$30\,\MSun$, the ISCO is high enough that lower-frequency parts of the
waveform contribute the most to the SNR.  For very high masses,
however, this basically cuts the waveform down to nothing.  In Initial
LIGO, the ISCO is completely buried in seismic noise for masses above
about $100\,\MSun$.  Thus, we must move the cutoff frequency up.  We
cannot push the cutoff far above ringdown, because the physical
waveform simply ceases to exist (see
Fig.~\ref{fig:StildesAndInitialPSD}).  It has been suggested that an
``effective ringdown'' (ERD) frequency $f_{\mathrm{ERD}} \equiv 1.07\,
f_{\mathrm{Ringdown}}$ is a useful upper limit~\cite{Pan2007}.  For
intermediate masses, we would like to interpolate somehow between
these two extremes of ISCO and ERD.  We suggest setting the cutoff
frequency to a weighted average of the two, where the weights are the
contributions to the SNR below the given frequency.  If we assume
coherent phasing between the template and the physical waveform, we
can simply take the amplitudes of the two waveforms.  Also, note that
the restricted SPA approximation for the amplitude is reasonable.
Thus, define
\begin{eqnarray}
  \label{eq:rhoISCO}
  \rho_{\mathrm{ISCO}}^{2} &\equiv & \int_{0}^{f_{\mathrm{ISCO}}}\,
  \frac{f^{-7/3}}{S_{n}(f)}\, d f\ , \\
  \label{eq:rhoERD}
  \rho_{\mathrm{ERD}}^{2} &\equiv & \int_{f_{\mathrm{ISCO}}}^{f_{\mathrm{ERD}}}\,
  \frac{f^{-7/3}}{S_{n}(f)}\, d f\ , \\
  \label{eq:rhoTOT}
  \rho_{\mathrm{tot}}^{2} &\equiv & \int_{0}^{f_{\mathrm{ERD}}}\,
  \frac{f^{-7/3}}{S_{n}(f)}\, d f\ , \\
  \label{eq:fCut}
  f_{\mathrm{cut}} &\equiv & \frac{f_{\mathrm{ISCO}}\ \rho_{\mathrm{ISCO}} +
    f_{\mathrm{ERD}}\ \rho_{\mathrm{ERD}}} {\rho_{\mathrm{tot}}}\ .
\end{eqnarray}
We have already dropped constant factors in the expressions for $\rho$
that will cancel out.

Note that these expressions only depend on the total mass by way of
the limits of integrations---which are very simple, known functions of
the mass---so these integrals could be done just once for a given
noise curve, storing the intermediate values.  When the cutoff needs
to be calculated, the cumulative integral could be evaluated at the
given ISCO and ringdown frequencies.  Hence, this would be a fast way
of calculating the cutoff, with no need to do the integrals each time
the cutoff is needed.

We can test this recommended frequency by comparing it to the optimal
cutoff frequency found by the amoeba search described in
Sec.~\ref{sec:Efficiency}.  For 3.5 pN templates in Initial LIGO, we
find that it is an excellent match to the optimal frequency.
Fig.~\ref{fig:FcRecomendations} shows these two values, along with
dark and light bands showing the regions in which changing $f_{c}$
results in a loss of overlap of 1\% and 3\%, respectively.  Of course,
the same figure shows that using the ERD recommendation would stay
within the 1\% error bounds.  Nonetheless, the close match between
this recommendation and the true optimum suggests that it is sound.
Thus, our final recommendation is to use the weighted-average
frequency cutoff throughout the entire mass range.  While our analysis
has been restricted to equal-mass systems, the cutoff frequency we
have defined here could be applied to unequal-mass systems as well.
It will be interesting to see how this cutoff fares in those
situations.

Similar results hold for Advanced LIGO, when using our recommended
template for each mass.  That is, in regions where 3.5 pN templates do
poorly (see Fig.~\ref{fig:ThreeParamOverlapSummaries}), the weighted
average is a poor predictor of the optimal cutoff frequency using
those templates, as shown in Fig.~\ref{fig:FcRecomendations}.
However, in those same regions---where pseudo-4.0 pN templates do
well---the weighted average is a good predictor of the optimal cutoff
frequency for 4.0 pN templates.  Thus, again, we recommend using the
weighted-average frequency cutoff throughout the entire mass range
with Advanced LIGO.

By prescribing a cutoff frequency, the search does not need to extend
over that parameter.  Similarly, by prescribing a post-Newtonian
order, we need use only one template for a given total mass.  On the
other hand, if these recommendations decrease the overlap found by too
much when using them compared to the overlap found by an unconstrained
search, it may be better to search the larger parameter space.  We can
evaluate the loss in overlap by comparing the results found using our
recommendations to the results found when searching over the set of
all three template families, and all masses, mass ratios, and cutoff
frequencies.  We have determined that this loss in overlap when using
our recommendations is always less than 0.0025 for Initial LIGO, and
less than 0.007 for Advanced LIGO.

\section{Conclusions}
\label{sec:Conclusions} %

We have compared high-accuracy NR waveforms for equal-mass binary
black holes from the Caltech--Cornell group to stationary phase
post-Newtonian waveforms.  We examined a number of factors that
influence the matches between the two, with the goal of optimizing the
matches and hence improving the efficiency of templated searches in
Initial and Advanced LIGO.  We first considered the effect of the
post-Newtonian order to which the phase evolution is taken, and found
that adding terms up to 3.5 pN or pseudo-4.0 pN to the currently-used
2.0 pN templates significantly improves the matches over a large range
of masses, as shown in Fig.~\ref{fig:ThreeParamOverlapSummaries}.  We
then studied the effect of varying the upper cutoff frequency of the
templates.  The frequency that achieves the optimal match is a
function of mass, and we find this function is well-approximated by an
average between ISCO and ERD, weighted by contribution to the SNR, as
shown in Fig.~\ref{fig:FcRecomendations}.  Finally, we allow the
symmetric mass ratio $\eta$ to range over unphysical values up to
$1.0$, and find that this dramatically improves matches, as shown in
Fig.~\ref{fig:PhysicalEta}.  Based on the results we recommend
adjusting the searches using \textit{TaylorF2} template waveforms by
going up to 3.5 pN or 4.0 pN over most of the mass range, integrating
up to our recommended cutoff, and allowing allowing $\eta$ to extend
up to 1.  For Initial LIGO, the overlaps obtained using these
parameters is always within 0.0025 of overlaps achievable by
optimizing over all three parameters.

In future work we plan to extend this analysis to unequal-mass and
spinning black-hole systems.  We have found that allowing unphysical
values of $\eta$ roughly doubles the size of the template bank, and we
also plan to study the impact of this on the false alarm rate.

\begin{ack}
  We thank Luisa Buchman, Yanbei Chen, Curt Cutler, Lisa Goggin, Larry
  Kidder, Luis Lehner, Lee Lindblom, Harald Pfeiffer, Mark Scheel and
  Kip Thorne for useful discussions.  This work was supported in part
  by grants from the Sherman Fairchild Foundation and by NSF grants
  PHY-0601459, PHY-0652995, and DMS-0553302 to Caltech.
\end{ack}



\section*{References}


\providecommand{\newblock}{}

\end{document}